\documentstyle[12pt]{article}
\begin{document}

\setlength{\unitlength}{1mm}

\textwidth 15.0 true cm

\textheight 22.0 true cm

\headheight 0 cm

\headsep 0 cm

\topmargin 0.4 true in

\oddsidemargin 0.25 true in

\setcounter{tocdepth}{3}
\thispagestyle{empty}

\noindent\hspace*{\fill}  FAU-TP3-98/18 \\



\begin{center}\begin{Large}\begin{bf} 

The Center-Symmetric Phase of QCD. \\  

\end{bf}\end{Large}\vspace{.75cm}

 \vspace{0.5cm}

F.Lenz \\

Institut f\"ur Theoretische Physik III \\

Universit\"at Erlangen--N\"urnberg \\

Staudtstra{\ss}e 7 \\

D-91058 Erlangen, Germany    

\end{center}

\vspace{1cm}

\begin{abstract} \noindent 

An investigation of the center symmetric phase of SU(2) QCD is
 presented.  The role of the center-symmetry, the dynamics of
Polyakov loops and the structure of Abelian monopoles are studied  within the axial gauge representation of QCD.  Realization
of the center symmetry is shown to result from non-perturbative gauge fixing
and concomitant confinement like properties emerging even at the perturbative level are
displayed. In an  analysis of the Polyakov loop dynamics, non-perturbative
gauge fixing is also shown to  
inevitably lead to singular gauge field configurations whose dynamics are briefly discussed.
\end{abstract}

\section{Introduction}  The center symmetry \cite{Suss79,McLerr81,Kuti81}
distinguishes the phases of Yang Mills theories. This symmetry is realized in the confining and
spontaneously broken in the deconfined phase with the Polyakov loops serving  as order
parameter. Unlike in studies of lattice QCD, the center symmetry has not been the subject of systematic analytical
investigations of QCD. For instance, perturbative approaches in general imply
a change of the underlying gauge symmetry from SU($N$) to U(1)$^{N^{2}-1}$ when
the coupling vanishes and thereby break the  Z$_N$ center symmetry. For the center symmetry to be preserved in the path integral formulation, the
Faddeev--Popov determinant  arising in the process of gauge fixing  cannot be treated
perturbatively. Likewise, for the center symmetry to be preserved in the  canonical
formalism, the Gauss law has to be resolved non-perturbatively. Only then the center
symmetry is guaranteed to appear as the correct residual gauge symmetry. The process
of non-perturbative gauge fixing unveils another fundamental and possibly far
reaching property of the formal structure of QCD. Unlike in QED, a global -- for all field
configurations valid -- elimination of redundant variables is possible in QCD only at the
expense of introducing coordinate singularities and thus of including singular gauge
field configurations in such gauge fixed formulations
\cite{SING78}. Formation of Gribov horizons \cite{GRIB78} or appearance of magnetic
monopoles \cite{MAND80,THOO81} represent two prominent examples of the
occurrence of singular field configurations as a result of non-perturbative gauge fixing.
 It is thus tempting to connect the realization of the center symmetry with the
emergence of singular field configurations and to identify these non-perturbative basic
structures as the origin of the characteristic properties of the confining phase of QCD. I
will present a study of the role of the center symmetry and the structure and dynamics
of monopole like singular field configurations in gauge fixed QCD. This discussion
summarizes the results of a series of investigations of QCD in axial gauge
\cite{Lenz2,Lenz3,LETH98,JALE98}.

 For the formulation of the center symmetry and
definition of the Polyakov loops, QCD has to be considered  in a geometry
where the system is of finite extent ($L$) in one direction ($x_{3}$), and  in
general of infinite
extent in the other directions. For finite temperature QCD, one has to choose the time direction to
be compact and the associated fields to be periodic (gauge fields) or
antiperiodic (quark fields). In the following discussion, the space-like 3-direction is assumed to be
compact. In this way, the center symmetry will appear as an ordinary symmetry
represented canonically by an operator which commutes with the
Hamiltonian. Such a standard interpretation of the center symmetry is not
possible for finite temperature QCD and  conceptual difficulties arise \cite{SMIL94}
concerning for instance the existence of domain walls. By
covariance, QCD at finite (spatial) extension is equivalent to finite
temperature QCD. 
By rotational invariance in the Euclidean, the value of the partition function of a system
with finite extension $L$ in 3 direction and $\beta$  in 0 direction is invariant under
the exchange of  these two extensions,
\begin{equation} Z\left(\beta,L\right)=Z\left(L,\beta\right) \ ,
\label{FE1}
\end{equation} provided standard boundary conditions in both time and 3 coordinate are imposed on the fields. As a consequence of (1), energy
density and pressure are related by
\begin{equation}
\epsilon\left(\beta,L\right)=-p\left(L,\beta\right) \ .
\label{FE2}
\end{equation} For a system of non-interacting particles this relation connects energy
density or pressure of the Stefan Boltzmann law with the corresponding quantities
measured in the Casimir effect.

In QCD, covariance also implies by Eq. (2) that at zero temperature a
confinement-deconfinement transition occurs when compressing the QCD vacuum (i.e.
decreasing $ L$). From lattice gauge calculations \cite{Kanayo} it can be inferred that
this transition occurs at a critical extension $L^{c} \approx 0.8$ fm in the absence of
quarks and at $L^{c} \approx 1.3$ fm when quarks are included. For extensions smaller
than
$L^c$, the energy density and pressure reach values which are typically 80 \% of the
corresponding  ``Casimir" energy and pressure. When compressing the system beyond
the typical length scales of strong interaction physics, correlation functions at
transverse momenta or energies $|p| \ll 1/L $ are dominated by the zero ``Matsubara
wave-numbers" in 3-direction and, as confirmed by lattice QCD calculations \cite{Reisz},
are  given by the dimensionally reduced QCD$_{2+1}$. 
\section{Center Symmetry}
The order parameter which characterizes the phases of QCD  is the vacuum expectation
value  of the trace of the Polyakov loop operator at finite temperature \cite{Svetitsky}
and correspondingly of the operator
\begin{equation} W \left(x_{\perp}\right) =  P\exp\Bigl\{ig \int_{0}^{L} dx^{3}  A_{3}
\left(x\right)  \Bigr\}
\label{FE3}
\end{equation} at finite extension ($x_{\perp} = (x_{0}, x_{1}, x_{2})$). I
will refer in the following also to $W$ as the Polyakov loop operator. Under
gauge transformations $U(x)$, $W\left(x_{\perp}\right)$ transforms as 
\begin{equation}
W(x_{\perp}) \rightarrow  U\left(x_{\perp},L\right)W(x_{\perp})
 U^{\dagger}\left(x_{\perp},0\right)\ .
\label{I31} 
\end{equation}
The coordinates  $x=(x_{\perp},0)$ and  $x=(x_{\perp},L)$ describe
identical points, and we require the periodicity properties imposed on the
field strengths not to change under gauge transformation. This is achieved if
$U$  satisfies
\begin{equation}
  \label{cs1}
 U\left(x_{\perp},L\right) = c_{U}\cdot U\left(x_{\perp},0\right)  
\end{equation}
with $c_{U}$ being an element of the center of the group. Thus gauge transformations can be classified according to the value of $c_{U}$ ($\pm 1$ in SU(2)). Therefore under gauge transformations
\begin{equation}
\mbox{tr}(W(x_{\perp})) \rightarrow 
\mbox{tr}(c_{U} W(x_{\perp}))
\stackrel{SU(2)}{=} \pm \mbox{tr}(W(x_{\perp})) .
\label{I31b} 
\end{equation} A simple example of an SU(2) gauge transformation $u_{-}$
with $c=-1$ is 
\begin{equation}
u_{-} = e^{-i\pi \vec{\tau} \hat{\psi}x_{3}/L}\quad, \quad c_{u_{-}}=-1 .
 \label{cs1a}
\end{equation}
with the arbitrary unit vector $\hat{\psi}$. Its effect on an arbitrary gauge field is
\begin{equation}
  \label{cs1b}
  A_{\mu}^{u_{-}} = e^{i\pi \vec{\tau} \hat{\psi} x_{3}/L}A_{\mu} 
e^{-i\pi \vec{\tau} \hat{\psi} x_{3}/L}-\frac{\pi}{gL}\vec{\tau} \hat{\psi}\delta_{\mu 3}. 
\end{equation}
This representative $u_{-}$ can
be used to generate any other gauge transformation changing the sign of $\mbox{tr}(W)$ by multiplication with a
strictly periodic ($c=1$) but otherwise arbitrary gauge transformation. The decomposition of $SU(2)$ gauge transformations into two classes according to $c=\pm 1$ implies a decomposition of each gauge orbit $\cal O$, into sub-orbits 
 $\cal O_{\pm}$ which are characterized by the sign of the Polyakov loop
\begin{equation}
  \label{cs8a}
 A(x)\, \epsilon \, {\cal O}_{\pm} \quad ,\quad \mbox{if}\quad \pm \mbox{tr}(W \left(x_{\perp}\right))\ge 0 .
 \end{equation}
Thus strictly speaking, the trace of the Polyakov loop is not a
gauge invariant quantity. Only $|\mbox{tr}(W(x_{\perp})|$ is invariant under all of
the gauge transformations. Furthermore, the spontaneous breakdown of the center symmetry as it supposedly happens at small extension
or high temperature is a breakdown of the underlying gauge symmetry. It
implies that the wave functional describing such a state is different for
gauge field configurations which belong to $\cal O_{+}$ and $\cal O_{-}$
respectively, and  which therefore are  connected by gauge transformations such as $u_{-}$ in Eq.(\ref{cs1a}). These considerations are also of relevance for understanding the structure of gauge fixed theories. Whenever gauge fixing is carried out exactly and with the help of strictly periodic gauge fixing transformations ($\Omega, c_{\Omega}=1$)
 the resulting formalism must contain the center symmetry 
\begin{equation}
\mbox{tr}(W(x_{\perp})) \rightarrow 
-\mbox{tr}(W(x_{\perp})) .
\label{I31a}
\end{equation} 
as residual gauge symmetry. In other words, gauge fixing via strictly periodic gauge transformations does not lead to a complete gauge fixing. Each gauge orbit is
represented by two gauge field configurations. This could be circumvented by
allowing for more general gauge fixing transformations which are periodic
only up to a center element, i.e. by including gauge fixing transformations with $c_{\Omega}=-1$. 

In the following we will carry out explicitly a gauge fixing procedure and
represent QCD in the axial gauge. As will be seen, the gauge fixing leading to axial gauge is incomplete in the above sense and will therefore exhibit the center symmetry as a residual gauge symmetry.  This gauge is of particular relevance for
properties related to the center symmetry, since the associated order
parameter, the Polyakov loops appear as elementary rather than composite
degrees of freedom. For carrying out the gauge fixing procedure the  following
transformation will be useful
\begin{equation}
  \label{cs3a} v_{-}=\Omega_{D}^{\dagger}\left(x_{\perp}\right) e^{-i\pi \tau_{3}
x_{3}/L}e^{i\pi\tau_{1}/2} \Omega_{D}\left(x_{\perp}\right)
\end{equation} where  $\Omega_{D}$ diagonalizes the Polyakov loop
\begin{equation}
  \label{cs3}
\Omega_{D}\left(x_{\perp}\right)W(x_{\perp})\Omega_{D}^{\dagger}\left(x_{\perp}\right)=
e^{i(gLa_{3}\left(x_{\perp}\right)+\pi)\tau_{3}} .  
\end{equation}  By including $\Omega_{D}$ in the definition of $v_{-}$, the color 3
direction and the color direction of the Polyakov loop coincide. 
\section{QCD in Axial Gauge} At this point we pass to a gauge fixed formulation by
applying the gauge fixing transformation    
\begin{equation}
  \Omega(x) = \Omega_{D}\left(x_{\perp}\right) (W^{\dagger}(x_{\perp}))^{x^{3}/L}
  \,
  P \exp\Bigl\{ig \int_{0}^{x^{3}} dz   A_{3}
    \left(x_{\perp},z\right)\Bigr\} .
\label{cs2} 
\end{equation}  in which the axial gauge is reached in 3 steps \cite{Lenz2}. In the
presence of the third factor only, the gauge transformation would eliminate $A_{3}$
completely. In order to preserve the periodic boundary conditions of the gauge fields
the second term reintroduces zero mode fields which in turn are diagonalized by
$\Omega_{D}$. Thus the gauge condition reads
\begin{equation} 
\label{za1}
\Omega\left(x\right)\Bigl(A_{3}(x_{\perp})+\frac{1}{ig}\partial_{3}\Bigr)\Omega^{\dagger}\left(x\right)=
 (a_{3} \left(x_{\perp}\right) +\frac{\pi}{gL}) \tau_{3}. 
\end{equation} 
By the gauge transformation, the 3 component of the gauge field is transformed
to zero apart from the eigenvalues of the Polyakov loops. The elementary rather than composite nature of the Polyakov
loop variables $a_{3}(x_{\perp})$  in axial gauge is
manifest. The gauge fixing transformation  $\Omega$  is periodic and
consequently field configurations which before gauge fixing are related by a gauge
transformation with $c=-1$ are not identified. Therefore center symmetry
transformations appear as residual symmetry transformations of the gauge fixed
theory. By construction, these symmetry transformations are the gauge fixed
transformations of Eq.(\ref{cs3a})
\begin{equation}
  \label{cs3aa} C=  \Omega\left(x\right)v_{-}\Omega^{\dagger}\left(x\right)   =e^{-i\pi
\tau_{3} x_{3}/L}e^{i\pi\tau_{1}/2} .
\end{equation} The effect of C on an arbitrary gauge field is most conveniently written
in a spherical color basis 
\begin{equation}
\Phi_{\mu}(x)= \frac{1}{\sqrt{2}}(A_{\mu}^{1}(x)+i A_{\mu}^{2}(x))e^{-i\pi x^{3}/L}
\label{gf12a} 
\end{equation} as 
\begin{equation}
  \label{gf12b} C: \quad a_{3} \rightarrow -a_{3}\quad ,\quad A_{\mu}^{3} \rightarrow
-A_{\mu}^{3}\quad ,\quad \Phi_{\mu} \rightarrow  \Phi_{\mu}^{\dagger} \quad ,\quad
(\mu \neq 3 ).
\end{equation} The center symmetry transformation $C$ acts as (Abelian) charge
conjugation with the ''photons'' described by the neutral fields $A_{\mu}^{3}(x),
a_{3}(x_{\perp}) $. For identification of the center symmetry with charge conjugation
symmetry, the shift in the definition of the Polyakov loop variables in Eq.(\ref{cs3}),
the rotation around the 1-axis in Eq. (\ref{cs3a}) as well as the shift in phase in the
definition of the charged fields (Eq.(\ref{gf12a})) have been introduced. As will be seen
shortly, this definitions will also simplify the description of the dynamics. The phase
change in Eq.(\ref{gf12a}) makes  the charged fields antiperiodic
\begin{equation}
\Phi_{\mu}(x_{\perp}, x^{3}=L) = - \Phi_{\mu}(x_{\perp}, x^{3}=0) . 
\label{cs3c} 
\end{equation}
If the center symmetry is realized  $ gL a_{3}({\bf x_{\perp}})$ has to be
distributed symmetrically around the origin. As will be seen below, other
variables exist in axial
gauge which can be used as order parameters of the
realization of the charge symmetry $C$. 

Apart from the discrete center symmetry transformation described by the charge
conjugation $C$, all other symmetries related to the gauge invariance have
been used to eliminate $A_{3}$. In such a case of a global, non-perturbative gauge fixing we have to expect, as argued above, singular field configurations
to emerge. In transforming to the axial gauge, diagonalization of the Polyakov loops ($\Omega_{D}$ in
Eq.(\ref{cs2}))  is the crucial step of the gauge fixing procedure, in which
such singular gauge
field configurations appear. This diagonalization can be viewed as choice of
coordinates in color space in which the color 3-direction is identified with
the direction of the Polyakov
loop. As is evident from Eq.(\ref{cs3}), this choice of coordinates becomes ambiguous if
$gLa_{3}(x_{\perp}^{N,S}) = \pm \pi$, i.e. if the Polyakov loop is in the center of the
group
\begin{equation} W(x_{\perp}^{N,S})= \pm {1\mkern-4mu\rm l} .
\label{za2}
\end{equation} This requirement determines a point on the group manifold ${\rm
S}^{3}$ and thus, for generic cases, fixes (locally) uniquely the position
$x_{\perp}^{N,S}$. At these points, the gauge transformed field 
\begin{equation}  A_{\mu}^{\prime} \left(x\right) = \Omega_{D}\left(x_{\perp}\right) 
A_{\mu} \left(x\right)\Omega_{D}^{\dagger}\left(x_{\perp}\right)
+s_{\mu}\left(x_{\perp}\right) \quad ,\quad \mu\neq 3 
\label{am5} 
\end{equation} with 
\begin{equation}  s_{\mu}\left(x_{\perp}\right) = \Omega_{D}\left(x_{\perp}\right) 
\frac{1}{ig}\partial _{\mu}\Omega_{D}^{\dagger}\left(x_{\perp}\right),  
\label{am5a} 
\end{equation} in general, is singular with $\Omega_{D}$.

All the elements are now available for writing down the central result of our investigations , the expression
for the axial gauge QCD partition function 
\begin{equation}
Z= \sum_{{\bf n}}Z_{{\bf n}} =
\sum_{{\bf n}}\int
D[a_{3}^{\bf n}]\int
\prod_{\mu\neq 3}D[A_{\mu}]
e^{-S[ A +s, a_{3}^{{\bf n}}]} .
\label{pm0a} 
\end{equation}
The integration variables, the unconstrained degrees of freedom, are the 3
components of the gauge field ($A_{\mu}(x), \mu \neq 3$) and the eigenvalues of the Polyakov loops. The integration over these eigenvalues has been decomposed according to the number
${\bf n}=(n_{N},n_{S})$ of north ($n_{N}$) and south ($n_{S}$) pole
singularities; i.e., the path integral in $Z_{{\bf n}}$ is performed over field
configurations in which the Polyakov loop passes $n_{N,S}$ times through north
and south pole respectively. For this decomposition to be meaningful,
regularization of the generating functional is required. The singular field
$s(x_{\perp})$
is determined by the Polyakov loop variables
\begin{equation}
  \label{pm0b}
  s = s \left[a_{3}^{{\bf n}}\right] .
\end{equation}
\section{Dynamics in Axial Gauge QCD.}
We will display the dynamical content of the above expression for the generating functional by discussing a hierarchy of approximations to $Z$ with increasing complexity.
\begin{enumerate} 
\item The QCD generating functional in the naive axial (or temporal) gauge is
  obtained if only the sector without singularities is kept and the dependence on the eigenvalues of the Polyakov loops is disregarded. As a consequence of these approximations, the generating functional becomes actually ill-defined as has been noticed by Schwinger 35 years ago \cite{Schwinger}. In definition of propagators certain ``i$\epsilon$'' prescriptions have to be applied. Due to the approximations, the center-symmetry is not present anymore.
\item Still, keeping the zero singularity sector only one might proceed by accounting
for the dependence of $Z$ on $a_{3}$. The simplest form of these dynamics
results, if these variables are treated as Gaussian variables, i.e. if the
non-flat measure
\begin{equation} d\left[a_{3}\right] = \prod_{y_{\perp}} \cos^{2}\left( gL
a_{3}(y_{\perp})/2 \right)\Theta \left( (\pi/g L)^{2}-a_{3}^{2}(y_{\perp}) \right)
 da_{3}\left(y_{\perp}\right)
\label{FE8}
\end{equation}
is replaced by the flat measure $ da_{3}$. In this way, one
effectively treats the Polyakov loop eigenvalues as the zero modes in QED. It
is therefore not surprising that the center-symmetry is lost again and Debye
screening like in QED \cite{Weiss} is obtained.

\item First characteristic properties of QCD are encountered if, still in the absence of singular 
field configurations, the non-flat measure of the Polyakov loop variables is
properly taken  into account. These properties will be the subject of the following
section. In particular, the perturbative phase reached in
this way will be seen to be center-symmetric. 
\item The role of singular field configurations in the $\bf{n}\neq {\bf 0}$
  sectors (cf.Eq.(\ref{pm0a})) is very poorly understood. In particular it has
  not been possible so far to identify those sectors which dominate the
  partition function nor has the dynamics of the quantum fluctuations around
  singular fields been studied systematically. Nevertheless, basic and well understood properties of
  QCD permit a certain indirect characterization of the dynamics in these
  sectors as will be discussed in the concluding section.  
\end{enumerate}
\subsection{Polyakov Loop Dynamics}
In this subsection we sketch the dynamics in the sector where no singularities are
present. Unlike in more standard approaches, the non-Gaussian
nature of the Polyakov loop variables $a_{3}(x_{\perp})$ is explicitly taken into
account and  the finite limit of integration associated with these variables is respected
\cite{LETH98}.We first consider the Polyakov loop dynamics in the absence
of coupling to the other degrees of freedom. The corresponding generating functional is,
in the Euclidean, given by
\begin{eqnarray} Z_{0} & = & \int d\left[a_{3}\right] 
\exp \left\{-1/2 \int d^{4}x(\partial_{\mu} a_{3}(x_{\perp}))^{2}\right\}
\label{I50} \\ & = & \int_{-\pi/2}^{\pi/2} \prod_{x_{\perp}}
d\tilde{a}_{3}\left(x_{\perp}
\right) \cos^{2}\tilde{a}_{3}\left(x_{\perp}\right)\exp \Bigl\{-\frac{2\ell} {g^{2}L}
\sum_{y_{\perp},\delta_{\perp}}(\tilde{a}_{3}( y_{\perp}+\delta_{\perp})-
\tilde{a}_{3}( y_{\perp}))^{2}\Bigr\} \ .
\nonumber
\end{eqnarray} Transverse space time has been discretized with $\ell$ and
$\delta_{\perp}$ denoting lattice spacing and lattice unit vectors respectively and the
Polyakov loop variables have been rescaled
\begin{displaymath}
\tilde{a}_{3}(x_{\perp}) = gLa_{3}(x_{\perp})/2 \ .
\end{displaymath} In the continuum limit,
\begin{equation}
\frac{\ell}{g^{2}L} \sim \frac{\ell}{L} \frac{1}{\ln \frac{\ell}{L}}
\rightarrow 0 \ ,
\label{I51}
\end{equation} and therefore the nearest neighbor interaction generated by the Abelian
field energy of the Polyakov loop variables is negligible. As a consequence, in the
absence of coupling to other degrees of freedom, Polyakov loops do not propagate,
\begin{equation}
\langle \Omega|T\left( a_{3} \left(x_{\perp}\right) a_{3} \left(0\right)
\right)|\Omega\rangle  \sim \Bigl(\frac{\ell}{g^{2}L}\Bigr) ^{x_{\perp}/\ell} \! \!
\rightarrow
\delta^{3}
\left(x_{\perp}\right) \ .
\label{I52}
\end{equation} Although the above procedure is similar to the strong coupling limit in
lattice gauge  theory, here a strong coupling approximation has not been invoked. In
the lattice dynamics of single links, the factor $1/g^{2}$ appears in the action and, as a
consequence,  continuum limit and strong coupling limit describe two different regimes
of the lattice theory. In the Polyakov loop dynamics on the other hand which is
controlled by the factor $\frac{\ell}{g^{2}L}$, strong coupling and continuum limit
coincide. Propagation of excitations induced by $a_{3} (x_{\perp})$ can consequently
only arise by coupling to the other microscopic degrees of freedom. Formally this
suggests the Polyakov loop variables
$a_{3} $ to be integrated out  by disregarding the contribution of the free $a_{3}$
action, but keeping the coupling to the other degrees of freedom. In this way,  the following effective action is obtained
\begin{equation} S_{\rm eff} \left[A_{\mu}\right] = S_{\rm YM} \left[A_{\mu},
A_{3}=0\right] + S_{\rm gf} \Bigl[ \int_{0}^{L} dz\, A_{\mu}^{3}\Bigr] +  M^{2}
 \int d^{4}x\,\, \Phi_{\mu}^{\dagger}(x) \Phi^{\mu}(x) .
\label{FE11}
\end{equation} 
The Polyakov loop variables have left their
signature in the  geometrical mass term of the charged gluons (cf. Eq.(\ref{gf12a}))
\begin{equation} M^{2}=(\pi^{2}/3-2)/L^{2}
\label{FE12}
\end{equation} and in the antiperiodic  boundary conditions  (Eq.\ref{cs3c}). The
neutral gluons remain massless and periodic. The antiperiodic boundary conditions
reflect the mean value of the Polyakov loop variables, the geometrical mass their
fluctuations; notice that in both of these corrections, the coupling constant has dropped
out. I emphasize that periodic boundary conditions for the gluon fields are imposed in
the representation (\ref{pm0a}) of the generating functional. The antiperiodic boundary
conditions in (\ref{cs3c}) describe the appearance of Aharonov-Bohm fluxes in the
elimination of the Polyakov loop variables. Periodic charged gluon fields may be used if
the differential operator
$\partial_{3}$ is replaced by
\begin{equation}
\partial_{3} \rightarrow \partial_{3}+\frac{i\pi}{2 L} \left[\tau_{3}\right. ,
  \qquad .
\label{FE13a}
\end{equation} As for a quantum mechanical particle on a circle, such a magnetic flux
is technically most easily  accounted for by an appropriate  change in boundary
conditions -- without changing the original periodicity requirements. With regard to
the rather unexpected physical consequences, the space-time independence of this flux
is important, since it induces global changes in the theory. These global changes are
missed if Polyakov loops  are treated as Gaussian variables. 

The role of the order parameter is taken over by the neutral color current in
3-direction $u\left(x_{\bot}\right)$ which is generated by the 3-gluon interaction 
\begin{equation} u \left(x_{\perp}\right) = i \int_{0}^{L} dx_{3}\,  \Phi^{\dagger}_{\mu}
\left(x\right)\stackrel{\leftrightarrow}{\partial}_{3}  \Phi^{\mu}
\left(x\right) \ .
\label{I54}
\end{equation} This composite field is odd under charge conjugation (cf.(\ref{gf12b}))
\begin{equation} C: \quad u(x_{\perp}) \rightarrow -u(x_{\perp}) . 
\label{FE14a} 
\end{equation} It determines the vacuum expectation value of the Polyakov loops
\begin{equation}
\langle \Omega| W \left(x_{\perp}\right)|\Omega \rangle \quad
\propto \quad \langle \Omega| u \left(x_{\perp}\right)|\Omega \rangle
\label{FE14}
\end{equation} and the corresponding correlation function
\begin{equation}
\langle \Omega| T\left[W \left(x_{\perp}\right) W \left(0\right)\right] |\Omega
\rangle  \quad  \propto \quad \langle \Omega| T\left[u
\left(x_{\perp}\right) u \left(0\right)\right] |\Omega \rangle \ 
\label{FE15}
\end{equation} which in turn yields the static quark-antiquark interaction energy
\cite{Svetitsky}. Up to an irrelevant factor we have after rotation to the Euclidean ($r= |
x^{E}_{\perp} |$)
\begin{equation} 
\label{FE17a}
\exp{\left\{-LV\left(r \right)\right\}}= \langle \Omega|T \left[u ( x^{E}_{\perp})
u\left(0\right)\right]| \Omega\rangle ,
\end{equation} i.e., the static quark-antiquark potential is given directly by (the
$a=b=3, \mu = \nu = 3$ component of)  the  vacuum polarization tensor $\Pi_{\mu
\nu}^{ab} $ and not by the zero mass propagator with corresponding self-energy
insertions as obtained in the standard Gaussian treatment. This remarkable
consequence of the ultralocality property (\ref{I52}) of the Polyakov loop variables
provides a direct connection between confinement and certain spectral properties of
gluonic states. If, as required in the center symmetric phase,  the vacuum expectation
value of the Polyakov loop operator vanishes and if the spectrum of states excited by
$u$ exhibits a gap $\Delta E$, Eq.(\ref{FE17a}) implies a linear rise in $V$ for large
separations
\begin{equation} V\left(r \right) \rightarrow  \sigma r = \Delta E  r/L \ .
\label{FE17b}
\end{equation} Thus in axial  gauge, confinement  is connected to a shift in the
spectrum of gluonic excitations to excitation energies
\begin{equation}
 E \geq  \sigma L \ 
\label{FE17c}
\end{equation} which diverges with the extension L becoming infinite. Comparison
with  the interaction energy of adjoint static charges suggests the negative charge
conjugation parity  (cf.Eq.(\ref{FE14a})) of the intermediate ''2-gluon'' states
contributing to $V$ in Eq.(\ref{FE17a})   to be the distinctive property which leads to
infinite excitation energies. 

The system described by the effective action (\ref{FE11}) exhibits remarkable
properties already at the perturbative level. Most importantly the center symmetry is
realized in the perturbative vacuum, i.e. in the ground state obtained by dropping all
the terms containing the coupling constant $g$. Geometrical mass (Eq.(\ref{FE12})) and
Aharonov-Bohm flux (Eq.(\ref{FE13a})) are not affected by such a perturbative
treatment. The perturbative ground-state is even under charge conjugation and the
expectation value of the Polyakov loop vanishes 
\begin{equation}
\langle\Omega_{\rm pt}| W \left(x_{\perp}\right)|\Omega_{\rm pt}
\rangle = 0 ,  \ 
\label{FE17}
\end{equation} indicating an infinite free energy of a static quark. Indeed perturbative
analysis of the correlation function (\ref{FE17a}) yields a linearly rising static
interaction energy. However the perturbative string tension decreases with increasing
extension
 ($ \propto L^{-2}$). The change from this value of the string tension to the physical
one together with the emergence of the proper QCD scale is beyond a perturbative
treatment also after elimination of the Polyakov loop variables. The perturbative
vacuum shares with the QCD vacuum certain properties also after including dynamical
quarks. In particular, application of perturbation theory shows the interaction energy
of static quarks to cease to rise indefinitely   and to be given at asymptotic separations
by the non-perturbative value of twice the mass of the dynamical quarks.        For
small distances, Coulomb-like behavior must emerge if the separation is small on the
scale of $\Lambda_{QCD}$ and small in comparison with the extension $L$.  This is
possible only, if the vacuum polarization tensor  possess an
 essential singularity at infinite momentum
\begin{equation}
\int d^{3}x\,  e^{i px}  \langle \Omega| T\left[u \left(x\right) u
\left(0\right)\right] |\Omega \rangle   \rightarrow e^{-\sqrt{g^{2}Lp/\pi}} .
\label{FE17d}
\end{equation} Obviously, finite order perturbation theory cannot yield such a
singularity; it however can be shown that, with increasing order in $g$, increasingly
high powers of $pL$ appear; two loop evaluation of the short distance behavior
indicates exponentiation. 

The perturbative phase with its signatures of confinement cannot be relevant for QCD at
extensions smaller than $L^c$. Not only do we expect the center symmetry to be
 broken at small extensions but also dimensional reduction to
 QCD$_{2+1}$ to happen. Due to the antiperiodic boundary conditions, charged gluons
decouple from the low-lying excitations if dimensional reduction takes place in the
center symmetric phase.   The small extension or high temperature limit of the center
symmetric phase is therefore QED$_{2+1}$. In order to reach the correct high
temperature phase, the deconfinement phase-transition arising when compressing the
QCD vacuum, must be accompanied by a change to periodic boundary conditions  and
simultaneously the geometrical mass must disappear. Connected with this change in the
charged gluon boundary condition is a change in Casimir energy density and pressure
which for non-interacting gluons (and neglecting the effects of the geometrical mass) is
given by
\begin{equation}
\Delta \epsilon=  - \pi^{2} /12 L^4 \ ,\qquad \Delta p = 3\Delta \epsilon \ . \label{FE18}
\end{equation} This estimate is of the order of magnitude of the change in the energy
density across the confinement-deconfinement transition when compressing the system,
\begin{equation}
\Delta \epsilon=  -0.45 /L^4 \ ,
\label{FE19}
\end{equation} deduced from the finite temperature lattice calculation of Ref.
\cite{Engels}.
\subsection{Axial Gauge Monopoles}
 In this concluding section I will characterize qualitatively the structure
 of the singular field configurations arising in the gauge fixing procedure
 and address some of the related dynamical issues (cf.\cite{JALE98}). For the following discussion
 it is convenient to identify, after a rotation to the Euclidean, time with
 $x_{3}$. In this way the singular fields (cf. Eq.(\ref{am5a})) are static
 magnetic fields.  A simple example of a singular
 field is that of a  
Dirac like monopole configuration  given by 
\begin{equation}
  {\bf s} \left({\bf x}\right)  
  = \frac{m}{2g} \left[-\frac{1+ \cos\theta}{2g r \sin\theta}
\hat{\varphi}\,\tau_{3}+\left((\hat{\varphi}+i m\hat{\theta})e^{\pm i
\varphi}\tau_{+}+\mbox{h.c.}\right)\right] \quad , \quad m=\pm 1. 
  \label{am18}
\end{equation} 
Here,  vectors denote (after rotation) the
spatial components (0,1,2). $\hat{\varphi}, \hat{\theta}$ are azimuthal and polar unit
vectors.
 The neutral component ($\propto \tau_{3}$) of the singular field ${\bf s}({\bf x})$ in
Eq.~(\ref{am18}) is  exactly the vector potential of a Dirac monopole \cite{DIRAC31} of
charge
$2\pi m/g$,with associated  magnetic field
\begin{equation}
  \label{b-Dirac}
  {\bf b}^{3} = \mathop{\rm rot} {\bf s}^{3}
  =  \frac{m}{2g} \frac{{\bf x}}{x^3} ,
\end{equation} and is accompanied by a singular charged field component ($\propto
\tau_{\pm}$). The singularity structure of the Dirac monopole configuration is not the
most general one. In addition to the longitudinal vector field ${\bf b}^{3}$, singular
transverse magnetic fields are also present whose strength is determined dynamically
and not quantized by topological requirements.
 
In 4-space, the transformed gauge fields are singular on straight lines parallel to the
time (3)-axis, and thus represent static singular magnetic fields.  The static nature of
the singularities is a trivial consequence of the static Polyakov loop which has been
selected for introducing coordinates in color space. North and south pole singularities
are distinguished by the value of the Polyakov loop (cf.Eq.(\ref{za2})). In addition to
poles, the field $ {\bf s}({\bf x})$ also exhibits (static) string like singularities
representing surfaces in 4-space. The singular neutral magnetic field ${\bf b}^{3}$, is
the central quantity in Abelian projected theories. The complete non-Abelian magnetic
field strength built from the inhomogeneous term of Eq.~(\ref{am5a}) and generated by
a gauge transformation of an everywhere regular gauge field cannot be singular and
vanishes,
\begin{equation} F_{i j}\left[s\right] =
\partial_{i}s_{j}-\partial_{j}s_{i}+ig\left[s_{i},s_{j}\right] = 0 ,
\label{za4}
\end{equation} since ${\bf s}$ is a pure gauge. \lq\lq Abelian'' magnetic monopoles
have vanishing magnetic field energy. Finally I mention the connection between axial
gauge monopoles and instantons. As is easily verified, the Polyakov loop of a single
instanton of size $\rho$ ($\rho \ll L$) is given by
\begin{equation} W({\bf x})= e^{i\pi \mbox{\boldmath$\scriptstyle\tau $}{\bf x} 
/\sqrt{{\bf x}^{2}+\rho^{2}}}
\label{Minst}
\end{equation} which shows that a single instanton contains a north and south pole
singularity at its center and at infinity respectively. More generally it can be shown
that the topological charge $\nu$ of a field configuration is given by the difference of
the net northern and southern charge
\begin{equation}
  \label{nu final}
  \nu = {\textstyle\frac12} \Bigl( \!\!
    \sum_{ i\atop W({\bf x}_i)=1} \!\!\!\!\! m_i 
    \, -  \!\!\!\!\! \sum_{ i\atop W({\bf x}_i)=-1}\!\!\!\!\!  m_i 
  \Bigr) .
\end{equation} 

On the basis of the connection between monopole formation and order parameter and 
using the link between monopoles and instantons, the dynamics of axial gauge
monopoles
 can, to some extent, be characterized. Condensation of monopoles is implied via
Eq.(\ref{nu final}) by results of the instanton liquid model \cite{SCSH96} and of lattice
QCD 
\cite{CGHN94} which suggest a finite instanton density in the QCD vacuum. However
since it also appears that instantons are not able to account for confinement
\cite{CADG78} monopole condensation itself does not appear to be  sufficient to induce the dual
Meissner effect. This is reminiscent of the difference in the response of a
plasma and a superconductor to a static external magnetic field. Obviously, instantons with their rigid correlation between north and
south pole singularities give rise to a very particular mode of condensation. Decoupling
of the singularities seems to be necessary for  generating the confined phase with a
symmetric distribution of north and south poles as  required by the center symmetry. 
Beyond the deconfinement transition condensation of axial gauge monopoles must be
expected to persist. With the center symmetry broken, the Polyakov loop is not
distributed symmetrically around the equator of $S^{3}$. It rather approaches more
and more either the north or the south pole with increasing temperature. An
expectation value
$W(x_{\perp})\neq \pm 1$ in the  infinite temperature limit is neither compatible with
the Stefan-Boltzmann law \cite{Lenz3} nor, as argued above, with the expected
dimensional reduction to 2+1 dimensional QCD. Thus, as the Polyakov loop approaches
one of the poles, the probability to pass through this pole and therefore the monopole
density must be expected to increase. On the other hand, for this increased density to be
compatible with perturbation theory and, in particular, not to lead to confinement, one
might expect poles and antipoles to compensate each other to a large extent. This would
be the case if poles and antipoles are strongly correlated with each other. We thus
expect the high-temperature phase to consist of a gas of magnetic dipoles and the
deconfinement-confinement transition to be be similar to the phase transition in the
2-dimensional XY model which occurs by vortex (monopole) unbinding. 
\section*{Acknowledgements} 
I thank M.Shifman and A.Vainshtein
for their hospitality and a vigorous discussion concerning the spontaneous
breakdown of the center symmetry. This work has been supported by the
Bundesministerium f\"ur Bildung, Wissenschaft, Forschung und Technologie. 
\section*{References}
\bibliographystyle{unsrt}

\end{document}